# Measuring Physical and Electrical Parameters in Free-Living Subjects: Motivating an Instrument to Characterize Analytes of Clinical Importance in Blood Samples

Barry K. Gilbert, Clifton R. Haider, Daniel J. Schwab, Gary S. Delp[1]

Special Purpose Processor Development Group, Mayo Clinic, Rochester, MN 55905, USA

## ABSTRACT

**Significance:** A path is described to increase the sensitivity and accuracy of body-worn devices used to monitor patient health. This path supports improved health management. A wavelength-choice algorithm developed at Mayo demonstrates that critical biochemical analytes can be assessed using accurate optical absorption curves over a wide range of wavelengths.

**Aim:** Combine the requirements for monitoring cardio/electrical, movement, activity, gait, tremor, and critical biochemical analytes including hemoglobin makeup in the context of body-worn sensors. Use the data needed to characterize clinically important analytes in blood samples to drive instrument requirements.

**Approach:** Using data and knowledge gained over previously separate research threads, some providing currently usable results from more than eighty years back, determine analyte characteristics needed to design sensitive and accurate multiuse measurement and recording units.

**Results:** Strategies for wavelength selection are detailed. Fine-grained, broad-spectrum measurement of multiple analytes' transmission, absorption, and anisotropic scattering are needed. Post-Beer-Lambert, using the propagation of error from small variations, and utility functions that include costs and systemic error sources, improved measurements can be performed.

**Conclusions:** The Mayo Double-Integrating Sphere Spectrophotometer (referred hereafter as MDISS), as described in the companion report [1], produces the data necessary for optimal component choice. These data can provide for robust enhancement of the sensitivity, cost, and accuracy of body-worn medical sensors.

## 1. INTRODUCTION

This paper describes a bio-analyte characterization process and the associated instrumentation that was developed to support that characterization. These instruments are used to provide the parameters to monitor clinically relevant medical data with body-worn devices. Improving body-worn clinical-grade health monitoring units has been a major end goal of our lab since the early 2000s. We began by implementing features on these units such as ECG and physical

---

activity, but always with the goal of incorporating additional measurement parameters into the units over time, e.g., blood oxygen saturation and carbon monoxide measurements, detection of methemoglobins, and other physiological parameters as feasible. This monitoring awaits appropriate reference data becoming available from laboratory-grade measurements. Discussion of measuring these additional analytes appears in [1].

Our intent in this manuscript is to highlight several separate research areas that coalesced in our laboratory over decades. These related threads led to our recognition of the need for a new state-of-art spectrophotometer that would yield the previously unavailable baseline data in support of the design of analyte-measuring untethered body-worn units. Although Section 3 of this paper may appear in part to be an historical review, that is not our primary intent; dedicated reviews are in the published literature. Rather, we wish to illustrate the way in which eight decades of prior work, much of it conducted in the authors' home institution, resulted in our most recent efforts, as described below. These fields have been continually active for decades; there are hundreds of references, some of which we cite herein.

The initial commercial development in the early 2000s of consumer-grade, non-analyte, body-worn units, and our development in the 2010s of clinical-grade, non-analyte, body-worn units, are described, followed by a brief review of Mayo Clinic's optically based analyte measurements originally made in the 1940s. Thereafter, our development of a high-performance research spectrophotometer system to collect data to support the design of battery-powered body-worn analyte measurement units is introduced. The companion report, [1], describes further spectrophotometer engineering details and presents example analyte measurement results from the MDISS.

## 2. INITIAL DEVELOPMENT OF CLINICAL-QUALITY-GRADE, BODY-WORN, PHYSIOLOGICAL MEASUREMENT UNITS

In the early 2000s, several consumer products companies began to market devices catering to the burgeoning field of self-help health-and wellbeing techniques, in particular, small, battery-operated monitoring devices, *e.g.*, a generic class of "step counters" and physical activity monitors [2-10]. These consumer units were not intended to be used in monitored clinical settings. However, we and our clinical colleagues at the Mayo Clinic needed similar units to measure the health and progress of patients, whose collected data would be of clinical grade. The outcome of these requirements was a multi-year project to develop high-quality, ruggedized, wearable sensing and recording devices, which could perform and record long-term motion tracking [11-18], as well as high fidelity monitoring of a free-living individual's electrocardiogram [19, 20]. Using miniaturized electronic components and microprocessors, and small high-energy-density batteries that became available in the first half of the 2000s, we also created ruggedized versions of these units for extreme-activity athletes and mountaineers [21] with several-week run times (24/7), as depicted in Figure 1. These studies were conducted under full written, informed consent according to Mayo Clinic Institutional Review Board (IRB) study IDs 11-006747, 12-001512 and 14-001445.

Before embarking on the development for clinical-quality wearable units, and to broaden our knowledge base, we began by purchasing, reviewing, and documenting the characteristics of several consumer self-help devices that were available in the open market [19] (as was also done by [2, 3]). Guided by these initial reviews of consumer-grade devices, and to ensure clinical-quality data, we incorporated features into our design such as: 1) Very high sampling rates of the measured analog signals, initially up to 400 samples/second, and later, up to 1000 samples/second, 8 or 12 bits/sample; 2) ECG waveform monitoring; 3) rigorous static and dynamic calibration of the accelerometers in every unit to NIST-traceable standards; 4) accelerometer slope and offset correction measurements on NIST traceable platforms, with data values stored to allow for post processed compensation for minor manufacturing differences (including a unique serial number and manufacturing date for each device for complete traceability); 5) autonomous multi-day operation without battery change or charging or any other intervention by the wearer; and 6) a stable time-of-day clock allowing the synchronization of data from multiple units worn by a single subject. Body-worn units designed, fabricated, tested, and deployed in this manner resulted in a reliable physiological measurement capability, in a small form factor, representing a set of potentially useful clinical tools for eventual monitoring of patients in their free-living environments.  We also demonstrated the ability to monitor patients, via short- and long-haul wireless and wired connections, from their home environments back to a medical center, where the collected raw data could be analyzed in near real-time [20].

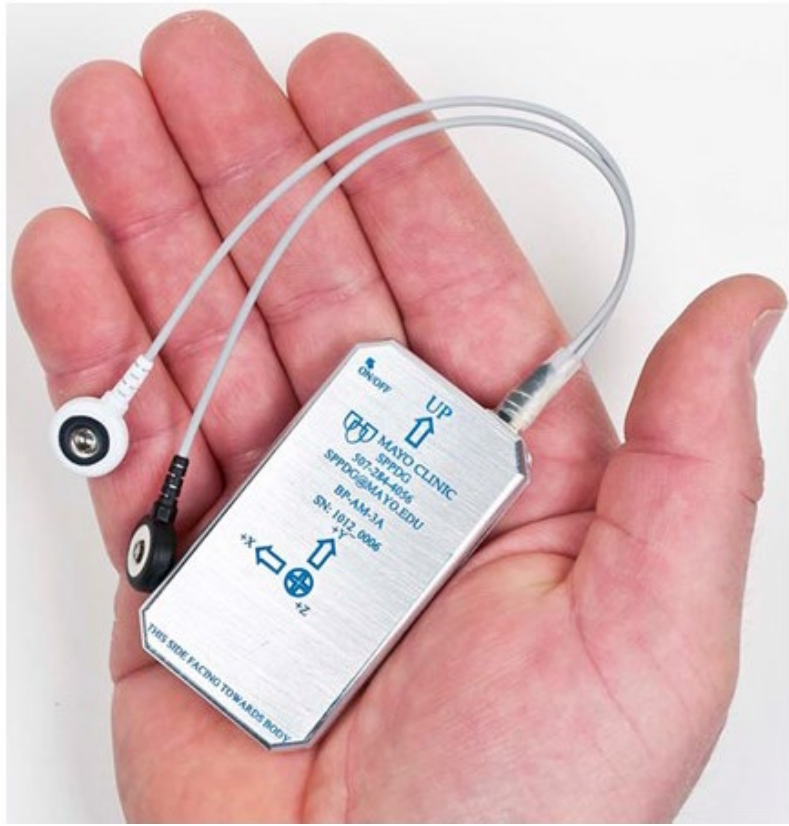

**Figure 1.** The small battery-powered body-worn units developed for continuous measurement of motion and ECG, worn by the mountain climbers on a Mayo-sponsored expedition to Mount Everest in 2012 [20]. The unit was 38.9 mm wide, 70.2 mm long, 8.9 mm thick; 3 3-axis accelerometers; 2 or 3-electrode, 400 samples/second ECG at 12-bit sampling resolution; 2-week run time. Our goal was to incorporate the analyte measurement capabilities into a physical form factor like these units (this figure also appears in U.S. Patent 8,849,387). (43333)[2]

---

[2] The (numbers) trailing the figure caption denote the figure's location in the SPPDG image archive.

## 3. EVOLUTION TO CLINICAL-QUALITY-GRADE, BODY-WORN BIOLOGICAL ANALYTE MEASUREMENT UNITS

The next request from our clinical colleagues was for an ability to measure and record, noninvasively and over a duration of days to a few weeks, the blood oxygen saturation levels in free-living patients (rather than in a hospital or clinic setting); we were also asked if it would be possible to measure the concentrations of other naturally occurring analytes as well.  The tiny, long-lasting body-worn units represented the target form factors into which our clinical collaborators asked us to incorporate these additional measurement modalities.

Regarding the measurement of blood oxygen saturation, we relied upon prior research in this field as a starting point.  We began by reviewing the significant body of work, beginning in 1935 and progressing steadily thereafter, on techniques for measuring blood oxygen saturation noninvasively.  This capability, using optical techniques based on two wavelengths of light, was first demonstrated in 1935 by Matthes [22], then extended by Milliken [23] and by Goldie [24] in the early 1940s.  Also, in the early 1940s, significant contributions to this field were made by Wood and colleagues [25-35].  However, the Wood team was unable to publish their results until 1947 and thereafter [36] because of wartime restrictions on the release of "sensitive" information since this work was conducted under the auspices of the U.S. Army Air Corps [though funded by Mayo Clinic as a contribution to the WW2 scientific efforts].  As with the work described in [22-24], the Wood earpiece oximeter employed two optical wavelengths, but it also incorporated a pressure-activated plunger to expel blood from the upper edge of the pinna (i.e., the upper portion of the outer ear) to achieve a hemoglobin-free tissue baseline that could be incorporated into the blood O2 calculations (Figure 2).  By present standards, the units were heavy and bulky, and had to be taped to the subject's head to provide mechanical support (Figure 3).  The Mayo-developed units were used in studies of G-induced loss of consciousness (G-LOC) in human subjects during World War II on a full-sized human centrifuge (partially visible in Figure 3) installed at the Mayo Clinic.  The earpiece oximeters continued in use without major changes until the early 1960s, in centrifuge studies of the Project Mercury astronaut couches.

The on-body oximetry technology continued to evolve.  The next significant advance, referred to as pulse oximetry (a variant on the original continuous oximetry) was first described in 1972 by Aoyagi and Kishi [37-39], with additional refinements in the subsequent decades.  The pulse oximetry approach effectively supplanted the original non-pulsatile approach in clinical implementations (see below).  Over the decades, commercial industry extended the implementation of pulse oximetry through hardware refinements using newer components and with algorithmic and software extensions to improve the usefulness and accuracy of the collected data, e.g., [40].  In 2000 Masimo introduced a technology approach referred to as Signal Extraction Technology (SET) [41], in which five proprietary algorithms were developed to remove the extraneous variability in the arterial oxygenation waveform caused by changes in the venous circulation, thereby providing a more accurate arterial oxygenation signal.

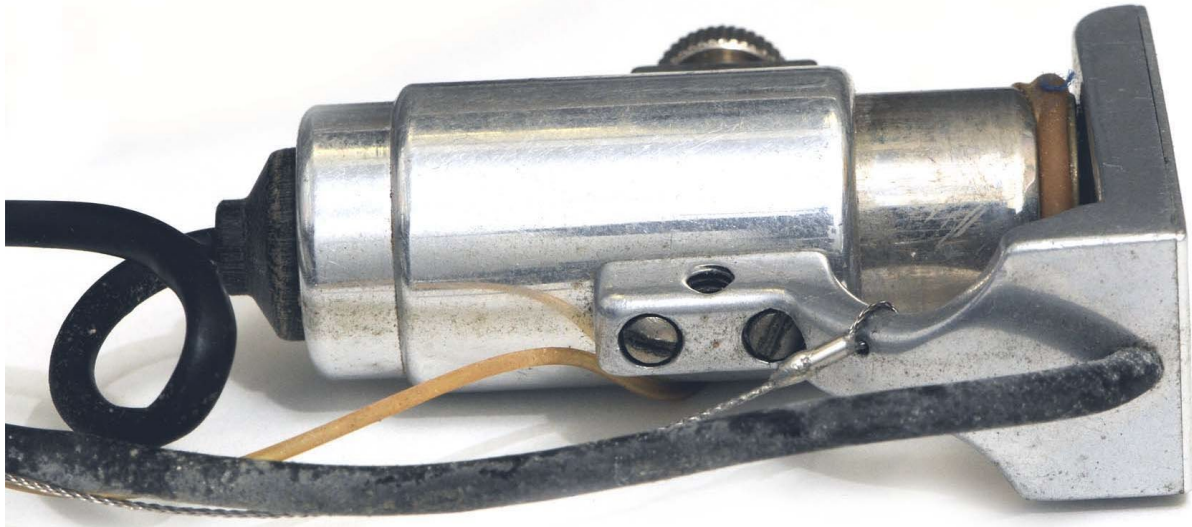

**Figure 2:** Earpiece oximeter developed at Mayo Clinic, illustrating light and plunger fully depressed against photodetector, ca. 1943. (42175)

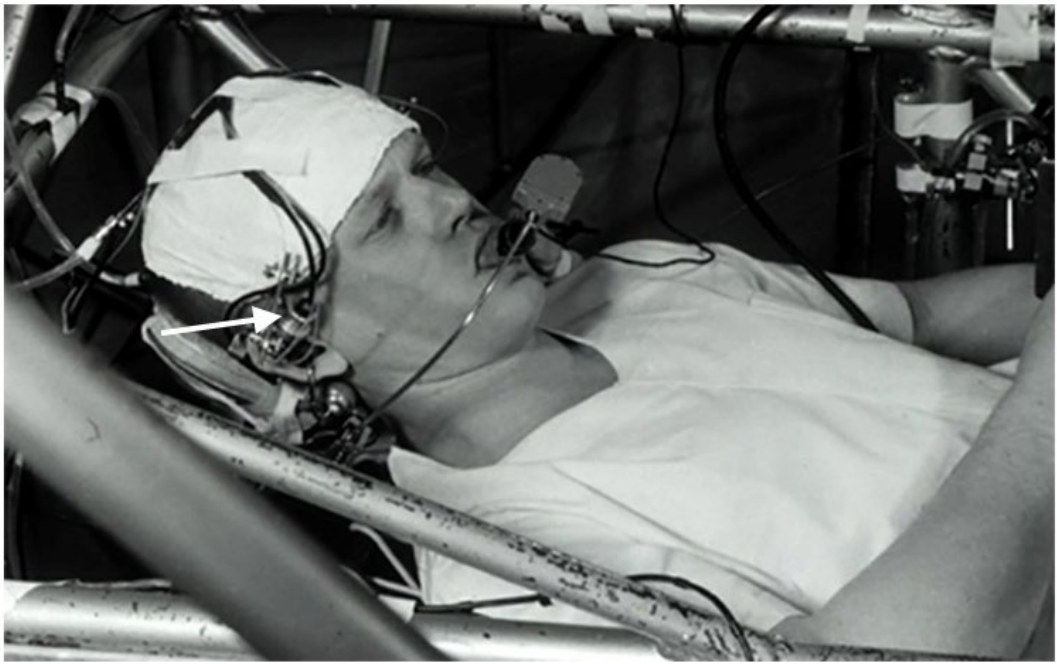

**Figure 3:** Volunteer in the cockpit of Mayo Clinic's human centrifuge, wearing earpiece oximeter on right ear, as indicated by white arrow, ca. 1962 (Photo courtesy of Don Hagland). (1954)[3]

---

[3] The (numbers) trailing the figure caption denote the figure's location in the SPPDG image archive. In this case, the phot was taken ca. 1962, and was the 1,954th entry in that archive.

Others have continued to document and refine the understanding of the physiological and optical processes underlying blood oximetry; see Severinghaus's excellent review of the early years of oximetry [39], and an exposition by Mannheimer of the optical physics and hemodynamics of the process [42]. By 2010, "finger-tip" oximeters, cable-powered by a desktop unit at the patient's bedside, were coming into use in hospital settings. These wired, clip-on devices are placed on the patient's index finger, and use "transmission oximetry", i.e., where light from two or more light-emitting diodes (LEDs) of different wavelengths passes through the finger, with the residual light then detected by one or more solid-state photodiodes.

To incorporate noninvasive analyte detection and measurement into our planned free-standing body-worn units we needed to identify optical components that would be compatible with the physical form factor constraints of the ECG/motion units. Our motion-and-ECG units were powered by small batteries. We viewed the analyte detection capability as an addition to the original baseline functionality. Thus, we needed to remain within the size and power constraints of those units, with the battery limitation being the most important. The consumer-grade LEDs used in the wired units require more power than we could support with small batteries. LEDs also have relatively wide emission bandwidths (20-70 nm full-width half-max [FWHM]) and uncertain center frequencies, which, as we later demonstrated, degrade the quality of data generated from them (discussed below). We turned to a family of small solid-state lasers, referred to as vertical cavity surface emitting lasers (VCSELs), which, in addition to their narrow-banded emission characteristics (FWHM optical bandwidths of 2-5 nm), were available over a wide range of optical center frequencies.

The first practical room temperature non-pulsatile (i.e., continuous-wave or CW) VCSEL was reported by Koyama et al. in 1988 [43]. Development of these tiny optical sources accelerated in the late 1990s and early 2000s by DARPA funding [44]. Our intent was to pair a small number of frequency-selected VCSELs with equally tiny solid-state photodetectors, either avalanche photodiodes (APDs) or P-I-N photodiodes (PIN diodes). APDs exhibit more signal gain than PIN diodes, but also have more intrinsic noise, so we concentrated on PIN diodes for our application. PIN diodes can be selected to cover a range of optical wavelengths. By combining narrow-spectrum VCSELs with PIN diodes having wide wavelength sensitivity, we could allow the light from VCSELS of different wavelengths to impinge on a single PIN diode. If in addition the light output from each VCSEL was modulated with a unique on-off keyed (OOK) pseudo-random pulsatile sequence (code division, Sig/Noise gain), and the PIN diode were integrated with an on-unit multi-channel correlator (i.e., in a small microcontroller or a custom correlator chip), accurate through-the-skin measurement of several analytes of clinical importance could be accomplished, simultaneously, in a small form factor, unlike in conventional pulse oximetry, where the measurements from the different LEDs must be made sequentially (thus introducing time skew between the two measured values in each pair).

The VCSELs' narrow emission lines indicated that considerable analyte measurement specificity could be achieved, far better than the LED-based wire-tethered fingertip pulse oximeters, based on Aoyogi's implementation [37-39], that entered hospital use in the early 2000s. However, to select the appropriate VCSEL wavelengths, we needed more frequency-accurate data than was identifiable in the published literature. It was this need for improved wavelength resolution, a larger wavelength measurement span, and additional wavelength measurement parameters that led to our decision to design and fabricate a spectrophotometer with extended functionality, as discussed below and in [1]. With this type of higher resolution and more comprehensive data available, it appeared feasible to design a wearable analyte sensor with considerably extended in vivo measurement capabilities compared to the then-commercial offerings. We undertook that effort. The goal of achieving an extended-capability autonomous on-body analyte measurement unit, underpinned by the historical evolution of such a capability as described above, represents the end goal of the entire sequence of projects described here and in [1].

Because we did not wish to be bound to through-the-finger transmission measurements, we investigated an alternate approach, referred to as "reflection oximetry", in which light is reflected from underlying bone, e.g., at the forehead, back to the photodetectors. The requirement for underlying bone also constrained placement options for the body-worn system. Therefore, we investigated a third approach, "scatter" oximetry, in which photons from the optical sources are directed into the skin and are sensed by one or more photodetectors placed several cm from the sources. Scattered photons entering the inputs of the photodetectors acquire and carry with them the optical information required to calculate the concentrations of the analytes of interest. Using scatter oximetry, the battery-powered measurement device can be placed on a wrist, on an arm, or on the torso, i.e., less intrusively than on a finger.

## 4. APPROACHES FOR SELECTING THE OPTIMUM MEASUREMENT WAVELENGTHS FOR ON-BODY OXIMETRY

Next, with this conceptual optical measurement chain sketched out, we needed a method to select the optimum wavelengths to measure analyte concentrations in vivo, so that the correct VCSELs could be incorporated into body-worn units. To address this problem one of us (CRH) developed a robust algorithm [45, 46] for the selection of optimal excitation wavelengths. Using the wavelengths selected by the algorithm allowed for the measurement of relative and absolute concentrations of a set of analytes in a sample.

The *Beer-Lambert* molar extinction coefficients of homogeneous materials can be measured for selected frequencies. Measurements in "the wild," however, need to incorporate many more factors, e.g., diffusion; heterogeneous paths; reflection; and the propagation of potential error from the inputs, through the measurement system, and continuing to a consideration of the variations and non-linearities of receivers. This system-level approach required more accurate and higher-resolution measurements of analyte characteristics, including diffusion, reflection, mean-free-path variation, florescence, phosphorescence, and various anisotropies. Our view of the importance of this information was informed by the guidance provided by [45, 46].

However, we did not have this data. Thus, we instituted a literature search for absorption curves for various forms of hemoglobin over a wide range of wavelengths. Such curves were first published in 1987 by Barker and Tremper [47], and again in 1989 by Tremper and Barker [48] (the authors credit Susan Manson, Biox/Ohmeda as their source of this data, as do many subsequent authors, which is widely accepted in the field), and are reproduced and duplicated in both the left and right panels of Figure 4; Figure 7 is from the same published sources. (Note: these curves might or might not be "correct" in some absolute sense, but they were the data available to us; "errors" in the shapes of these curves would of course have deleterious effects on the results that we present below but were unknowable.) Next, we searched the published literature for a selection of wavelengths that would yield the most accurate measures of the concentrations of four forms of hemoglobin (COHb, MetHb, OxyHb, and DeoxyHb). Lamego *et al.* [40] describes *eight* "optimum" wavelengths (610, 620, 630, 655, 700, 720, 800 and 905 nm). The wavelengths in the rightmost panel of Figure 4, illustrated as vertical lines, are similar though not identical to those documented in Lamego *et al*.

The left panel of that figure depicts four wavelengths calculated from the algorithm of [45, 46] also provides a means of computing a quality factor, referred to as the "condition number" for a selected set of wavelengths, in which a lower number is "better". A condition number for the eight wavelengths in the right panel of Figure 4, taken from [40] as the outcome of a literature search, was calculated as 31.2, whereas the condition number for the four wavelengths in the left panel of that figure was calculated to be 19.8, i.e., a better condition number for the use of fewer wavelengths. The algorithm of [45, 46] teaches the following conditions, all of which must be satisfied simultaneously as best possible: 1) each selected measurement wavelength should be maximally separated from other wavelengths; 2) the measured curves should be maximally separated in the vertical, or extinction coefficient, axis; 3) wavelengths should not be selected in regions where an extinction curve is "steep", since slight differences in the shapes of the curves and/or in their left-to-right or vertical positions due to component variations or other differences can introduce errors in the final resulting measurements (this constraint is often difficult to obey). In the rightmost panel, the wavelengths 610, 620, and 630 nm are "too close together". The wavelengths at 660, 730, 805 and 905 nm are "good" choices, according to our algorithm, although the overall condition number is degraded by the wavelengths at 610, 620, and 630 nm.

The four wavelengths selected by our algorithm, in the left panel of Figure 4, are in partial violation of the ground rules described above; the wavelengths are constrained to some extent by the shapes of the curves themselves. In addition, there is a fifth curve, labeled "Penalty", which we incorporated to account for the fact that some VCSELs are more difficult and costly to manufacture than others; the shape of this curve changes over time as manufacturing processes evolve [courtesy of M. Hibbs-Brenner, Vixar Corporation, *ca*. 2010]. Were the penalty curve not incorporated into the calculations, the selection of wavelengths would have been somewhat different, and might yield results which obey the above-described guidelines more closely.

Finally, in Figure 4, note the addition of small, bell-shaped curves at the bottom of the rightmost panel, which presents the wavelengths published by a commercial vendor [40]. These bell-shaped curves illustrate the wavelength spread generated by LEDs on either side of their center frequencies, which, as commented above, can be 20-70 nm (FWHM) wide. The LEDs thus generate wavelength components which overlap one another, thereby degrading the measured signal-to-noise ratios at the photodetector circuitry, because each light source will be carrying "information" that optimally would be out-of-band for those light sources. VCSELs generating FWHM wavelengths only a few nm in width will not create or will at least minimize these artifactual components, thereby motivating their use in place of LEDs.

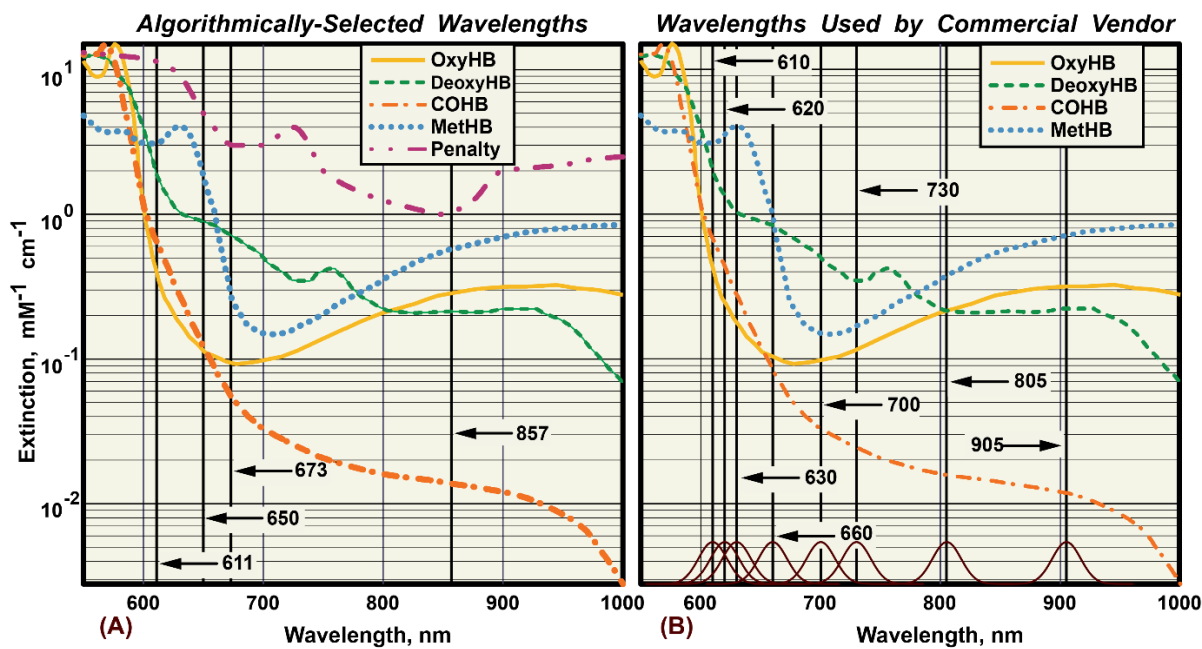

**Figure 4**: Oxyhemoglobin, Deoxyhemoglobin, Carboxyhemoglobin, and methemoglobin extinction coefficients as a function of wavelength (four algorithmically selected maximally linearly independent wavelengths in the left panel, versus eight wavelengths selected by a commercial vendor in the right panel). Narrow-band VCSEL sources are assumed, though commercial vendors typically employ LEDs [40], whose wider bandwidths and overlaps are illustrated by the bell-shaped curves at the bottom of the rightmost panel. Condition numbers: left panel: 19.8; right panel: 31.2. (42003)

As is clear from this chart, the optical wavelengths at which meaningful hemoglobin measurements can be performed are in the range of approximately 450 nm to 1000 nm. As will be noted below, several other analytes can also be measured using a similar implementation, but the wavelength ranges needed to be extended down to 200 nm and up to 2000 nm.

## 5. THE NEED FOR, AND OUR APPROACH TO OBTAINING, AN IMPROVED SPECTROPHOTOMETER

The results described above demonstrated that with good continuous optical absorption data over a broad wavelength span, the algorithm described here could select the optimum wavelengths, and hence the correct VCSELs, to detect one or more naturally occurring analytes present in sufficient concentration in the blood of an individual as measured by an autonomous battery-operated (i.e., untethered!) body-worn unit. We also recognized that the more optical parameters from laboratory samples that we had, the more specifically we could select the optimum VCSEL wavelengths. To gather the requisite data, we needed a spectrophotometer capable of measuring many optical parameters simultaneously, in a wide variety of physiological samples (e.g., whole blood, plasma, lymph, etc.).

A review of available commercial spectrophotometers confirmed that such a machine did not exist. The available models for purchase measured only two parameters, *i.e.*, transmission and backscatter, and were lacking most of the desired operational features. We also conducted a literature search against the possibility that other research groups had developed very capable spectrophotometers, from which they might have published high accuracy absorption curves of one or more naturally occurring analytes. We discovered developments, both theoretically and in some cases through the implementation of actual hardware [49-52] that addressed some of the parameters that appeared to us to be important, but none that were sensitive to as many parameters as our studies indicated might be physically implementable and clinically relevant. In several of those previously published papers, descriptions of the characteristics of those machines were limited. Therefore, we elected to design and fabricate an optical instrument capable of measuring more of these parameters, over a wide spectral range, in very short durations. Our intent was to create a set of baseline data that would enable the next step, that of designing the desired autonomous body-worn units. Following a brief introduction to the spectrophotometer developed here, we will return to a discussion of the value of the extended measurement parameters that we believed to be important.

### 5.1. The Mayo Double-Integrating Sphere Spectrophotometer (MDISS): Lessons-Learned Regarding the Selection of Optical Wavelengths for Body-Worn Analyte Measurement Units

Only a summary of the features of the MDISS is presented here; a complete description appears in [1]. From a functional perspective, we consider the MDISS to be a multi-generational successor to the Beckman DU spectrophotometer [53-55]; also, we wished to extend the capabilities of the experimental machines previously referenced with a combination of quantitative functionality features including: a broad wavelength measurement span (190-2750 nm); high wavelength accuracy and high wavelength precision (FWHM on the order of 1 nm); fine wavelength resolution (0.1-5 nm wavelength step sizes); high amplitude sensitivity; a rapid-throughput automated measurement capability; simultaneous acquisition of diffuse reflected (DR), diffuse transmitted (DT), and unscattered transmitted (UT) energy; input energy monitoring for high accuracy energy values on a pulse by pulse basis (with options for fluorescent and opto-acoustic acquisition); and the flexibility to accommodate a wide variety sample holder types and volumes, including high-pressure (up to 30 atmosphere [450 psi]) cells for measurements of blood oxygen saturation characteristics in hyperbaric environments). We also attempted to mitigate potential sources of measurement error of parameters, as will be noted below. Figure 5 is a photo of an early version of the unit.

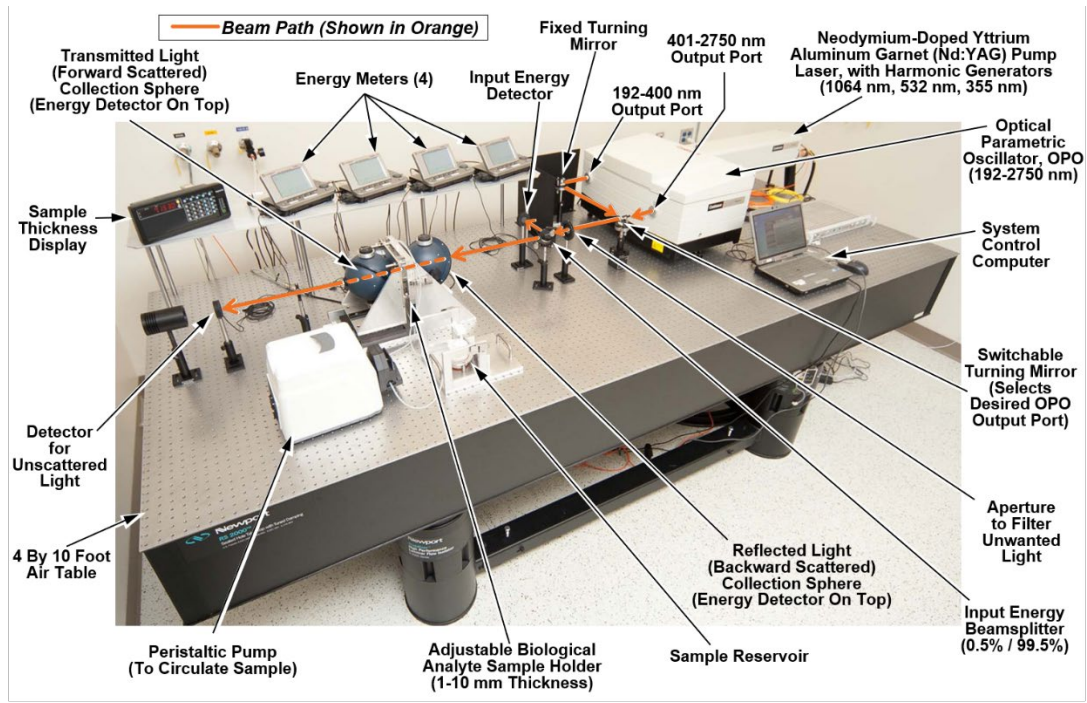

**Figure 5:** Double-integrating sphere spectrophotometer system (narrow-linewidth pump laser coupled to a tunable optical parametric oscillator, allowing precise optical characterization of biological analytes over a 192-2750 nm wavelength span. (44414)

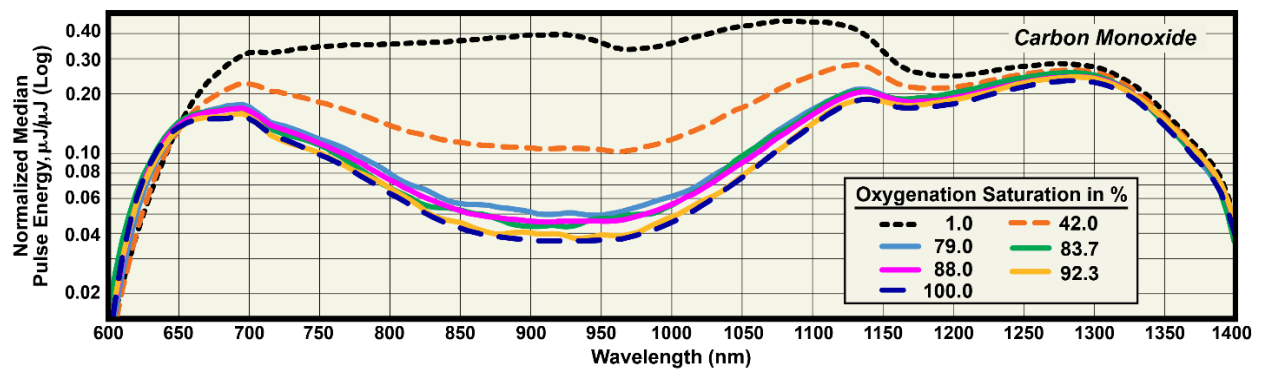

**Figure** 6: A set of spectroscopic transmission curves measured with the MDISS at each of six oxygen saturation levels, at wavelength separations of 2-nm, over a range of 800 nm. The desaturation gas in this example was carbon monoxide (note the isosbestic point at 650 nm, rather than at 800 nm if the desaturation gas had been carbon dioxide). (46584)

## 5.2. MDISS Design Tradeoffs

The design of the MDISS system required a recognition of the various cost tradeoffs in terms of money, time (development time and actual test time), and specific performance parameters. A single example of a specific performance parameter trade-off was prioritizing wavelength span rather than absolute energy sensitivity of the detectors. This selection was accepted to address the numerous unknowns regarding the analyte characteristics, since we had no prior knowledge of those subregions in the entire wavelength span in which a given analyte would display maximum amplitude differences. This requirement for sensors with wide optical bandwidths justified the use of pyroelectric sensors, which have broad wavelengths spanning 190–12,000 nm, and the use of an optical parametric oscillator (OPO) which produced energy from 192 to 2750 nm. If it had been determined that, even with the broadband capability, there was a need for more sensitivity, then alternate detectors could have been implemented, albeit at the expense of reduced bandwidth. As was discovered during the blood hemoglobin and oxygenation characterizations with different desaturation gases, most of the variations that we needed to detect occurred in the 600-1200 nm span.

In a mutually reinforcing fashion, the underlying theory was subsequently corroborated by the measured results from the MDISS when it became operational. Without the machine, which was designed and constructed under the assumption that the guidance gleaned from [45, 46] was correct, many if not most of the results demonstrated by the machine to are valid, and necessary for the design of accurate body-worn analyte measurement units, would have been unavailable.

Noninvasive optical sensing can be used with other clinically important physiological and biochemical variables besides hemoglobin species. As one example, Figure 7 displays absorption curves for blood glucose, blood protein, and blood lipids in the near-IR range of 1350 nm to 1850 nm. As in Figure 4, these waveforms were identified and employed following a search of the open literature. An absorption curve for water is provided for comparison. With the appropriate sensing wavelengths as determined by the algorithm and indicated by the vertical lines in this figure, detection and measurements of glucose, lipids, proteins, and even a measure of either total body water or central blood volume could be made.

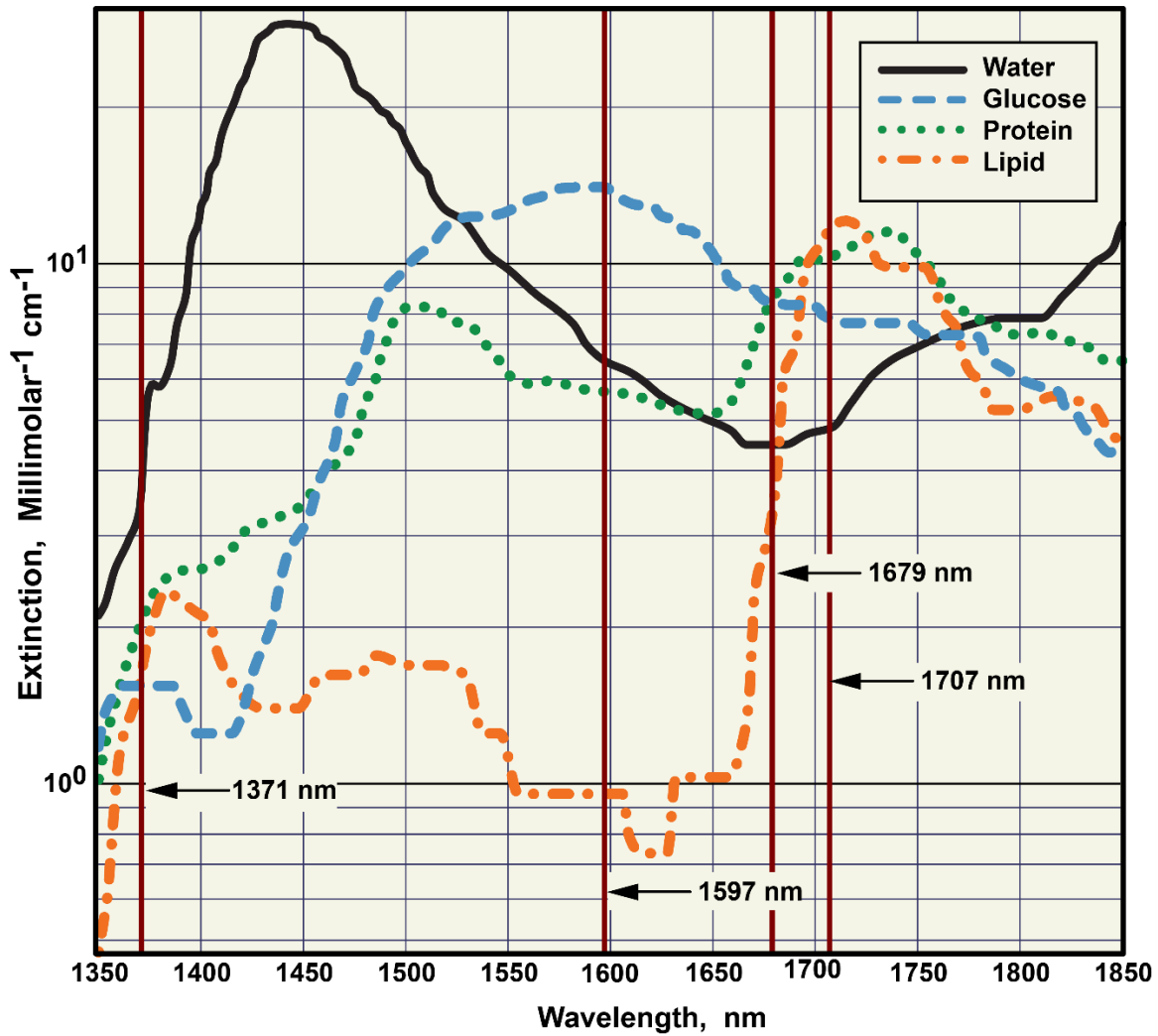

**Figure 7:** Water, protein, and glucose extinction coefficients as a function of wavelength, assuming ideal 1-nm bandwidth light sources using maximally independent wavelengths. Condition number: 10.9. (41964)

## 6. THE LONG-TERM GOAL OF THIS EFFORT

The long-term goal of this multiple-step project was the development of a sufficient technical base so that small, battery-powered, microcontroller-enhanced body-worn units could be designed to measure and record, with high accuracy and in real-time, blood oxygen saturation, blood levels of carbon monoxide, and concentrations of other clinically relevant analytes. The first developmental stage was the MDISS instrument, which was to yield the type of clinically actionable data described above. A next step, described in [55], was the prototyping of the battery-powered electrical and optical components to conduct these measurements continually, in a sufficiently small form factor that they could be incorporated into versions of the body-worn units depicted in Figure 1, that could be fielded into the clinical practice. In addition, although we did not pursue this path, commercial versions of the MDISS system could be used as research instruments; or, reduced-wavelength-span and/or lower cost versions targeted for a subset of analytes could be used for higher throughput routine clinical diagnostic purposes.

# 7. SUMMARY

We have described the evolution of consumer-grade body-worn physiological measurement units. We then introduced an evolutionary thread from early work in the development of *research-grade* body-worn blood oxygen saturation units conducted in the first half of the 1940s. Next, we reviewed our recognition in the 2010s of the requirements for higher-quality laboratory measurements of the optical characteristics of medically relevant blood analytes (*e.g.*, oxygen saturation-versus-wavelength behavior of critical blood analytes). We extended that line of investigation with the design of a spectrophotometer, the MDISS, with the needed sensitivity and specificity to help us gather new optically based blood characterization data [1]. We also initiated efforts to use the results reported here and in [1] to develop a new-technology body worn unit that operates on a somewhat different principle from classical pulse oximetry. Although body-worn units were our major end goal, we have also underscored the benefits of using optical analyte data over a wide wavelength range with high wavelength resolution. An instrument with capabilities beyond those available is required; these new data support a future effort to simplify and optimize body-worn monitors.

*Disclosures*

The authors declare no conflicts of interest. Although various patents cover various aspects of this work, none of these patents are licensed for gain or profit.

*Acknowledgements*

We wish to acknowledge the years-long contributions of the following individuals to this project: Charles Burfield, Anthony Ebert, Theresa Funk, Nicholas Klitzke, Steven Polzer, Jason Prairie, and Steven Schuster; and Drs. Franklyn Cockerill, Kendall Cradic, Graham Cameron, E. Rolland Dickson, Stefan Grebe, Ravinder Singh, and Nathan Harff. The clinical materials used in the study were produced by the Mayo Clinic Department of Laboratory Medicine and Pathology and processed by them for us using their standard clinical laboratory tools.